\def\rfr#1{eq. (\ref{#1})}

\def\virg#1{``#1''}
\def\bb#1#2#3{\bibitem[\protect\citeauthoryear{#1}{#2}]{#3}}

\def\dert#1#2{\frac{{{d}}{#1}}{{{d}}{#2}}}              

\def\dert#1#2{\frac{{{d}}{#1}}{{{d}}{#2}}}              

\def\bar{\begin{eqnarray}}
\def\ear{\end{eqnarray}}
\def\eqi{\begin{equation}}
\def\eqf{\end{equation}}
\def\eqia{\begin{eqnarray}}
\def\eqfa{\end{eqnarray}}
\def\rp#1#2{\frac{#1}{#2}}

\def\lb#1{\label{#1}}

\def\bds#1{\boldsymbol{#1}}

\documentclass{aastex}
\usepackage{spr-astr-addons}
\usepackage{url}
\usepackage{amsmath,amsthm,amscd,amssymb}
\usepackage{latexsym}
\usepackage{graphicx,epsfig}

\RequirePackage{color}

\newcommand{\emaila}{lorenzo.iorio@libero.it}

\begin{document}

\title{On the MOND External Field Effect in the Solar System }
\shorttitle{On the MOND External Field Effect in the Solar System }
\shortauthors{L. Iorio}

\author{Lorenzo Iorio\altaffilmark{1} }
\affil{INFN-Sezione di Pisa. Permanent address for correspondence: Viale Unit\`{a} di Italia 68, 70125, Bari (BA), Italy.}

\email{\emaila}

\begin{abstract}
In the framework of the MOdified Newtonian Dynamics (MOND), the internal dynamics of a gravitating system $s$ embedded in a larger one $S$ is affected by the external background field $\bds E$ of $S$ even if it is constant and uniform, thus implying a violation of the Strong Equivalence Principle: it is the so-called External Field Effect (EFE). In the case of the solar system, $E$ would be $A_{\rm cen}\approx 10^{-10}$ m s$^{-2}$ because of its motion through the Milky Way: it is orders of magnitude smaller than the main Newtonian monopole terms for the planets. We address here the following questions in a purely phenomenological manner:  are the Sun's planets affected by an EFE as large as $10^{-10}$ m s$^{-2}$? Can it be assumed that  its effect is negligible for them because of its relatively small size? Does $\bds E$ induce vanishing net orbital effects  because of its constancy over typical solar system's planetary orbital periods? It turns out that a constant and uniform acceleration, treated perturbatively, does induce non-vanishing long-period orbital effects on the longitude of the pericenter $\varpi$ of a test particle. In the case of the inner planets of the solar system and with $E\approx 10^{-10}$ m s$^{-2}$, they are $4-6$ orders of magnitude larger than the present-day upper bounds on the non-standard perihelion precessions $\Delta\dot\varpi$ recently obtained with by E.V. Pitjeva with the EPM ephemerides in the Solar System Barycentric frame. The upper limits on the components of $\bds E$ are   $E_x\leq 1\times 10^{-15}$ m s$^{-2}$,  $E_y\leq 2\times 10^{-16}$ m s$^{-2}$, $E_z\leq 3\times 10^{-14}$ m s$^{-2}$.
This result is in agreement with the violation of the Strong Equivalence Principle by MOND.
Our analysis also holds  for any other exotic modification of the current laws of gravity yielding a constant and uniform extra-acceleration. If and when other corrections $\Delta\dot\varpi$ to the usual perihelion precessions will be independently estimated with different ephemerides  it will be possible to repeat such a test.
 \end{abstract}


\section{Introduction}
The MOdified Newtonian Dynamics (MOND) \citep{Mil83a,Mil83b,Mil83c} is a non-linear theory of gravity which undergoes departures from the standard Newtonian inverse-square law at a characteristic acceleration scale; it was proposed to explain certain features of the motion of ordinary electromagnetically detectable matter in galaxies and of galaxies in galactic clusters  without resorting to exotic forms of still undetected dark matter. MOND, among other things,  does not fulfill the Strong Equivalence Principle in the sense that the internal dynamics of a gravitating system $s$ of bodies does depend on the external background gravitational field $\bds E$ of a larger system $S$ in which $s$ is embedded, even if $\bds E$ is constant and uniform \citep{Mil83a,joint,Mil86}; it is the so-called External Field Effect (EFE).

Its effects have mainly been  treated in a number of ways in the context of galactic dynamics \citep{San02,Fam07,Angus08,Wu}, while they have received some attention in the case of the Oort cloud in the solar system \citep{Mil83a,Mil86}. It is just the case to recall that it moves through the Milky Way at about 8.5 kpc from its center; its gravitational attraction at the solar system's location can be evaluated from the magnitude of its centrifugal acceleration \citep{Mil83b} $A_{\rm cen}\approx A_0$, where \citep{Bege} $A_0=1.27\times 10^{-10}$ m s$^{-2}$ is the MOND characteristic acceleration scale. Thus, in principle, such an EFE should affect the inner dynamics of the solar system. Concerning such an issue,
a researcher active in MOND writes: \virg{MOND breaks down the Strong Equivalence Principle. This means that the acceleration of solar system's bodies depends indeed on the background gravitational field and not only on the tidal field. As shown by Milgrom, even if the
external field was constant (and the tidal force vanishes), the internal acceleration would depend on the external field. Claiming that $A_{\rm cen}$ is irrelevant is only valid if the field equation were linear.} He/she also adds that \virg{for trans-Neptunian objects and planets, one can ignore the $A_{\rm cen}$.}
Another researcher working on MOND tries to go in deeper details  by  writing: \virg{For the main planets, the acceleration
is much larger than $A_0$ (the order of magnitude of the EFE), and the effect is negligible [...]
The EFE maintains a constant direction in the planet revolution, and its effect cancels out. }
Such statements are likely expressions of a widely diffuse belief about EFE in solar system.

In this paper we deal with statements like those previously quoted by showing, in a purely phenomenological way, that they are, in fact, not valid for the motion of the major bodies of the solar system referred to the Solar System Barycenter (SSB) frame usually adopted for planetary data reduction.
In particular, we will address the following points
\begin{itemize}
  \item Does a constant and uniform acceleration $\bds E$ having a generic direction $\bds{\hat{l}}$ and a magnitude of the order of $E=A_{\rm cen}\approx 10^{-10}$ m s$^{-2}$  affect the motion of the major bodies of the solar system?
  \item Are the effects of such an acceleration negligible for the Sun's planets, although they are, in principle, present?
  \item Since $\bds E$ is constant and uniform, are its effects canceled out over planetary revolutions?
\end{itemize}
We will demonstrate that
\begin{itemize}
  \item A constant and uniform acceleration does induce non-zero long-period, i.e. averaged over one orbital revolution, effects on the Keplerian orbital elements of a planet
  \item By assuming $E\approx A_0$, the resulting perihelion precessions of the inner planets are $4-6$ orders of magnitude larger than the present-day limits on the recently estimated non-standard perihelion rates \citep{Pit05}.
\end{itemize}
\section{The impact of EFE on the dynamics of the inner planets}
Such an external field $\bds E$, which would not be aligned with the internal Newtonian attraction $\bds N$, but it would be directed along a generic direction $\bds{\hat{l}}$ fixed during one orbital revolution, can be modelled as
\eqi \bds E = E_x\ \bds i + E_y\ \bds j + E_z\ \bds k,\lb{acc}\eqf
where $E_x,E_y,E_z$ are assumed constant and uniform, and $\bds i,\bds j,\bds k$ are the unit vectors of the usual SSB frame used to describe the planetary motions.  The constraints we will obtain are, thus, of dynamical origin and model-independent.

\subsection{The perihelion precession induced by a constant and uniform perturbing acceleration}
From an observational point of view,   the astronomer E. V. \citet{Pit05} has recently estimated, in a least-square sense, corrections $\Delta\dot\varpi$ to the standard Newtonian/Einsteinian averaged precessions of the longitudes of the  perihelia of the inner planets of the solar system, shown in Table \ref{chebolas},
\begin{table}
\caption{Estimated corrections $\Delta\dot\varpi$, in units of $10^{-4}$ arcsec cty$^{-1}$ (1 arcsec cty$^{-1} = 1.5\times 10^{-15}$ s$^{-1}$),  to the standard Newton/Einstein perihelion precessions
of the inner planets according to Table 3 of \citet{Pit05} (Mercury,
Earth, Mars). The result for Venus has been obtained by recently processing radiometric
data from Magellan spacecraft (E.V. Pitjeva, private communication, 2008). The reference frame used is the usual Solar System Barycentric one.\label{chebolas}
}
\centering
\begin{tabular}{|c|c|c|c|}
\hline
\multicolumn{1}{|c|}{Mercury} & \multicolumn{1}{|c|}{Venus} & \multicolumn{1}{|c|}{Earth} & \multicolumn{1}{|c|}{Mars}  \\
\cline{1-4}
%
%
$-36\pm 50$ & $-4\pm 5$ & $-2\pm 4$ & $1\pm 5$\\
\hline
\end{tabular}
\end{table}
by fitting almost one century of planetary observations of several kinds with the dynamical force models of the EPM ephemerides; since they do not include the action of any exotic modified model of gravity, such corrections are, in principle, suitable to constrain the unmodelled action of such a putative EFE accounting for its direct action on the inner planets themselves.
A preliminary, back-on-the-envelope assessment of the impact of an additional acceleration $E$ having the magnitude of $A_0$ on the planetary perihelia can be done by simply taking the ratio of   $E$ to the average speed of a planet along its orbit, given by $v\approx na$, where $n=\sqrt{GM/a^3}$ is the un-perturbed Keplerian mean motion. For the Earth $v\approx 3\times 10^4$ m s$^{-1}$, so that $\dot\varpi\approx E/na = 3$ arcsec cty$^{-1}$, which is about 4 orders of magnitude larger than the present-day uncertainty in the upper bound on any Earth's anomalous perihelion precession (Table \ref{chebolas}). However, an accurate calculation must ultimately be done, also because
our simple order-of-magnitude evaluation cannot tell us if, in fact, non-vanishing net secular perihelion precessions really occur under the action of a constant and uniform acceleration $\bds E$.

In order to precisely calculate the effects of \rfr{acc} on the orbit of a planet, let us project it onto the radial ($\bds{\hat{r}}$), transverse ($\bds{\hat{t}}$) and normal ($\bds{\hat{n}}$) directions of the co-moving frame picked out by the three unit vectors  \citep{Mont}
\eqi \bds{\hat{r}} =\left(
       \begin{array}{c}
          \cos\Omega\cos u\ -\cos I\sin\Omega\sin u\\
          \sin\Omega\cos u + \cos I\cos\Omega\sin u\\
         \sin I\sin u \\
       \end{array}
     \right)
\eqf
 \eqi \bds{\hat{t}} =\left(
       \begin{array}{c}
         -\sin u\cos\Omega-\cos I\sin\Omega\cos u \\
         -\sin\Omega\sin u+\cos I\cos\Omega\cos u \\
         \sin I\cos u \\
       \end{array}
     \right)
\eqf
 \eqi \bds{\hat{n}} =\left(
       \begin{array}{c}
          \sin I\sin\Omega \\
         -\sin I\cos\Omega \\
         \cos I\\
       \end{array}
     \right)
\eqf
%
  %
  %
%
where $\Omega$ is the longitude of the ascending node which yields the position of the line of the nodes, i.e. the intersection of the orbital plane with the mean ecliptic at the epoch (J2000), with respect to the reference $x$ axis pointing toward the Aries point $\Upsilon$, $I$ is the inclination of the orbital plane to the mean ecliptic  and $u=\omega+f$ is the argument of latitude defined as the sum of the argument of the pericentre $\omega$, which fixes the position of the pericentre with respect to the line of the nodes, and the true anomaly $f$ reckoning the position of the test particle from the pericentre.
Thus,  it is possible to obtain the radial, transverse and normal components of the perturbing acceleration as
\begin{eqnarray}
  E_r &\equiv & \bds E\bds\cdot \bds{\hat{r}}, \\ \nonumber \\
  E_t & \equiv & \bds E\bds\cdot \bds{\hat{t}}, \\ \nonumber \\
  E_n & \equiv & \bds E\bds\cdot \bds{\hat{n}};
\end{eqnarray}
a straightforward calculation shows that they are linear combinations of $E_x,E_y,E_z$ with coefficients proportional to harmonic functions whose arguments are lineal combinations of $u$, $\Omega$ and $I$.   $E_r,E_t$ and $E_n$
 must be inserted  into the right-hand-side of the Gauss equations \citep{Ber} of the variations of the Keplerian orbital elements
{\tiny{
\begin{eqnarray}\lb{Gauss}
\dert a t & = & \rp{2}{n\sqrt{1-e^2}} \left[e E_r\sin f +E_{t}\left(\rp{p}{r}\right)\right],\lb{gaus_a}\\  \nonumber \\
\dert e t  & = & \rp{\sqrt{1-e^2}}{na}\left\{E_r\sin f + E_{t}\left[\cos f + \rp{1}{e}\left(1 - \rp{r}{a}\right)\right]\right\},\lb{gaus_e}\\ \nonumber \\
\dert I t & = & \rp{1}{na\sqrt{1-e^2}}E_n\left(\rp{r}{a}\right)\cos u,\\  \nonumber \\
\dert\Omega t & = & \rp{1}{na\sin I\sqrt{1-e^2}}E_n\left(\rp{r}{a}\right)\sin u,\lb{gaus_O}\\   \nonumber \\
\dert\omega t & = &\rp{\sqrt{1-e^2}}{nae}\left[-E_r\cos f + E_{t}\left(1+\rp{r}{p}\right)\sin f\right]-\cos I\dert\Omega t,\lb{gaus_o}\\  \nonumber \\
\dert {\mathcal{M}} t & = & n - \rp{2}{na} E_r\left(\rp{r}{a}\right) -\sqrt{1-e^2}\left(\dert\omega t + \cos I \dert\Omega t\right),\lb{gaus_M}
\end{eqnarray}
}}
where  $e$ and ${\mathcal{M}}$ are  the eccentricity and the mean anomaly of the orbit of the test particle, respectively,  and $p=a(1-e^2)$ is the semi-latus rectum.
By evaluating them onto the un-perturbed Keplerian ellipse
\eqi r = \rp{a(1-e^2)}{1+e\cos f}\eqf    and averaging\footnote{We used $\dot u=\dot f$ because over one orbital revolution the pericentre $\omega$ can be assumed constant.} them over one orbital period $P_{\rm b}$ of the test particle  by means of
\eqi \rp{dt}{P_{\rm b}} = \rp{(1-e^2)^{3/2}}{2\pi(1+e\cos f)^2}df,\eqf
it is possible to obtain the long-period effects induced by \rfr{acc}.

In order to make contact with the latest observational determinations, we are interested in the averaged rate of the longitude of the pericenter $\varpi\equiv \omega+\Omega$;
its variational equation is
{\tiny{
\eqi\dert\varpi t = \rp{\sqrt{1-e^2}}{nae}\left[-E_r\cos f + E_{t}\left(1+\rp{r}{p}\right)\sin f\right]+2\sin^2\left(\rp{I}{2}\right)\dert\Omega t .\lb{varpidot}\eqf
}}  %
%
%
%
%
%
 %
 %
%
%
%
%
According to \rfr{varpidot}, it turns out that, after lengthy calculations, the averaged rate of $\varpi$ consists of a long-period signal given by a linear combination of $E_x.E_y,E_z$ with coefficients ${\mathcal{C}}_j,\ j=x,y,z$ of the form
\eqi {\mathcal{C}}_j = \rp{1}{na}\sum_k F_{jk}(e){\cos\xi_{jk}},\ j=x,y,z\eqf
where $F_{jk}(e)$ are complicated functions of the eccentricity and $\xi_{jk}$ are linear combinations of $\varpi,\Omega$ and $I$. In principle,
they are time-varying according to
\begin{eqnarray}
  \varpi &=& \varpi_0+\dot\varpi t, \\
  \Omega &=& \Omega_0 + \dot\Omega t, \\
  I &=& I_0 + \dot I t;
\end{eqnarray}
from a practical point of view, since their secular rates $\dot\varpi,\dot\Omega,\dot I$ are quite smaller, we can assume $\varpi\approx\varpi_0,\Omega\approx\Omega_0,I\approx I_0$ in computing $\cos\xi_k$, where $\varpi_0,\Omega_0,I_0$ are their values at a given reference epoch (J2000); see on the WEB http://ssd.jpl.nasa.gov/txt/p$\_$elem$\_$t1.txt.
\section{Constraints on EFE}
Using the estimated extra-rates of Venus, Earth and Mars of Table \ref{chebolas}, it is possible to  write down a non-homogenous linear system of three equations in the three unknowns $E_x,E_y,E_z$  by equating the corrections $\Delta\dot\varpi$  for the aforementioned planets to their predicted perihelion precessions due to $\bds E$. It is
{\small{
\eqi \Delta\dot\varpi_p = {\mathcal{C}}_{px} E_x + {\mathcal{C}}_{py} E_y + {\mathcal{C}}_{pz} E_z,\ p = {\rm Venus,\ Earth,\ Mars}.\lb{curlo}\eqf
}}
The values of the coefficients ${\mathcal{C}}_{pj}$ are in Table \ref{coffi}; they have been computed by using  for $\varpi,\Omega,I$ the values at the reference epoch (J2000)  by JPL, NASA (http://ssd.jpl.nasa.gov/txt/p$\_$elem$\_$t1.txt), but it can be shown that Table \ref{coffi} does not appreciably change if we vary them within the error bars of the EPM\footnote{In fact, EPM values of the planetary orbital elements the at (J2000) are not publicly available.} ephemerides retrievable, e.g., from Table 3 by \citet{PitSha}.
\begin{table}
\caption{ Computed values of the coefficients ${\mathcal{C}}_{j}$, in s m$^{-1}$, for Venus, Earth and Mars. For $\varpi,\Omega,I$ entering $\cos\xi_{jk}$ the values at the reference epoch (J2000) have been used (http://ssd.jpl.nasa.gov/txt/p$\_$elem$\_$t1.txt).\label{coffi}
}
\centering
\begin{tabular}{|c|c|c|c|}
\hline
\multicolumn{1}{|c|}{}  & \multicolumn{1}{|c|}{${\mathcal{C}}_x$} & \multicolumn{1}{|c|}{${\mathcal{C}}_y$} & \multicolumn{1}{|c|}{${\mathcal{C}}_z$}  \\
 \cline{1-4}
Venus & $0.0023$ & $-0.0052$ & $-0.0002$\\
Earth & $-0.0005$ & $-0.0019$ & $5\times 10^{-10}$ \\
Mars & $-0.0005$ & $9\times 10^{-6}$ &  $-1\times 10^{-10}$ \\
\hline
\end{tabular}
\end{table}
As a result, we have  for the Cartesian components of the perturbing acceleration $\bds E$ the figures quoted in Table \ref{acci}.
{\small{
\begin{table}
\caption{ Cartesian components of the perturbing acceleration $\bds E$, in m s$^{-2}$, from the means and standard deviations of the solutions of the eight different systems of equations of the type of \rfr{curlo} obtained by combining the maximum and minimum values of $\Delta\dot\varpi$ within their ranges according to Table \ref{chebolas} ($+++;++-;+-+;-++;+--;--+;-+-;---$).
\label{acci}
}
\centering
\begin{tabular}{|c|c|c|}
\hline
\multicolumn{1}{|c|}{$E_x\ (\times 10^{-15})$} & \multicolumn{1}{|c|}{$E_y \ (\times 10^{-16})$} & \multicolumn{1}{|c|}{$E_z  \ (\times 10^{-14} )$} \\
 \cline{1-3}
$-0.3\pm 1$ & $2\pm 5$  &  $-0.6\pm 3$ \\
\hline
\end{tabular}
\end{table}
}}
We have obtained them as follows. First, we have considered the eight different systems of the type of \rfr{curlo} obtained by taking the maximum and the minimum values of $\Delta\dot\varpi$ for Venus, Earth and Mars according to their best estimates and uncertainties of Table \ref{chebolas}; then, after solving them, we computed the means and the standard deviations of  the three sets of eight solutions for $E_x,E_y,E_z$.
The upper bounds on $E_x,E_y,E_z$ are $E_x\leq 1\times 10^{-15}$ m s$^{-2}$,  $E_y\leq 5\times 10^{-16}$ m s$^{-2}$, $E_z\leq 3\times 10^{-14}$ m s$^{-2}$,  respectively.  The total acceleration is
\eqi E = (0.6\pm 3)\times 10^{-14}\ {\rm m\ s}^{-2}.\lb{costr}\eqf
Such a constrain, which is a conservative one since we linearly added the uncertainties in $E_x,E_y,E_z$ to obtain
\eqi \delta E\leq \rp{\left|E_x\right|\delta E_x + \left|E_y\right|\delta E_y  + \left|E_z\right|\delta E_z  }{E} ,\eqf is four orders of magnitude smaller than $A_0$. This substantially confirms our preliminary, non-analytical evaluation.
\section{Interpretation of the results obtained}
\begin{figure}
\includegraphics[width=\columnwidth]{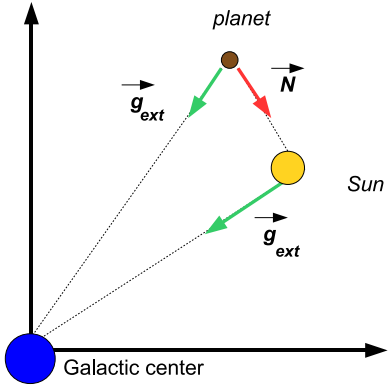}
 \caption{\label{GCframe} Sun-planet system $s$ embedded in an external system $S$ given by the Milky Way: Galactocentric frame. In it $\bds N$ (red arrow) is the Newtonian gravitational attraction of the planet by the Sun (the Newtonian gravitational attraction of the Sun by the planet is not depicted), while $\bds E\equiv\bds g_{\rm ext}$ (light green arrow) is the  gravitational attraction of the planet by the Galaxy whose magnitude amounts to about $E\approx 10^{-10}$ m s$^{-2}$ (in principle, it should also account for the gravitational attraction by other sources external to the Galaxy itself as the galaxy M31 Andromeda, etc.: however, their acceleration is of the order of $\approx 0.01 A_0$).   }
\end{figure}

How could we interpret the results obtained here? In fact, there is no contradiction with the findings in \citep{joint,Mil86} because they are valid in an inertial Galactocentric frame, for which the boundary condition at infinity yielding \eqi \bds\nabla\varphi\rightarrow -\bds E\ {\rm for\ } \bds r\rightarrow\infty, \eqf where $\varphi$ is the solution of the modified Poisson equation for the gravitational potential in the non-relativistic theory of MOND proposed by \citep{joint}, holds. See Figure \ref{GCframe} in which we have indicated the Newtonian monopole term with $\bds N$ and the Galactic attraction   $\bds E$ with $\bds g_{\rm ext}$.

\begin{figure}
\includegraphics[width=\columnwidth]{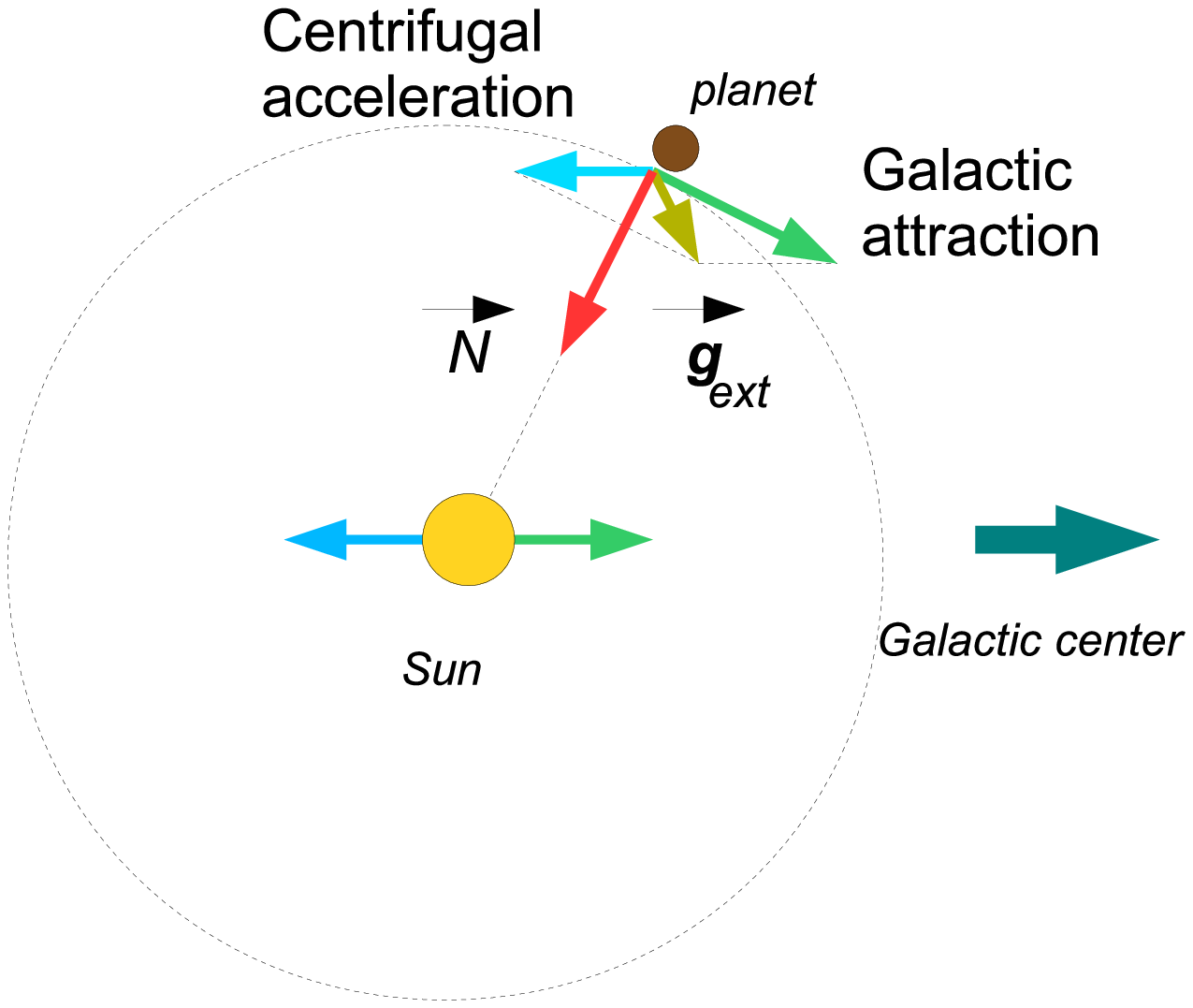}
 \caption{\label{SSBframe} Sun-planet system $s$ embedded in an external system $S$ given by the Milky Way: heliocentric frame. In it $\bds N$ is the Newtonian gravitational attraction of the planet by the Sun (red arrow), while $\bds E$ is given by the gravitational tidal forces $T$ (olive green arrow) exerted on the planet by the Galaxy whose magnitude amounts to about $T\lesssim 10^{-14}$ m s$^{-2}$; they account for both the uniform centrifugal accelerations due to the motion of Solar System through the Galaxy (light blue arrow) and the non-uniform gravitational attraction by the Galaxy throughout the extension of the Solar System (light green arrow).}
\end{figure}

It is not so in the SSB frame used for usual solar system's planetary data reduction because of the centrifugal acceleration which (almost) exactly cancels the Galactic gravitational attraction throughout SSB leaving just a small net tidal residual $T$, if any, whose magnitude can be easily evaluated as $T\lesssim 10^{-14}$ m s$^{-2}$. See Figure \ref{SSBframe}.

In other words, the outcome of the gravitational experiment represented by the planetary motion does depend on the velocities of the inertial frames in which it is studied, which are GC (at rest) and SSB (freely falling in the Milky Way). After all, this should not surprise too much, since MOND does, indeed, violate the Strong Equivalence Principle.
\section{Conclusions}
In this paper we dealt with the problem of the External Field Effect in solar system in the framework of the MOdified Newtonian Dynamics. In particular, we analyzed the common belief that, in principle, also the internal motions of the Sun's planets are affected, in MOND, by a constant and uniform acceleration $\bds E$ having the same magnitude of the centrifugal acceleration of the solar system's revolution around the Galactic center $A_{\rm cen}\approx 10^{-10}$ m s$^{-2}$, and directed along a fixed direction $\bds{\hat{l}}$ in space.
It is believed that its magnitude is too small to affect the motions of the major bodies of the solar system, and that its net effects vanish because of its constancy during  one planetary orbital revolution.
We found, instead, that such a kind of perturbing acceleration does, in fact,  produce non-vanishing net long-period, i.e. averaged over one orbital revolution, orbital effects on the motion of  a test particle. Among them, of particular relevance are the perihelion precessions because it is possible to compare them to the corrections to the standard Newtonian/Einsteinian perihelion rates of the inner planets recently estimated with the EPM ephemerides in the SSB frame. As a result, we obtained upper bounds on the components of $\bds E$ by finding $E_x\leq 1\times 10^{-15}$ m s$^{-2}$,  $E_y\leq 2\times 10^{-16}$ m s$^{-2}$, $E_z\leq 3\times 10^{-14}$ m s$^{-2}$, with $E\leq 3\times 10^{-14}$ m s$^{-2}$, i.e. $4-6$ orders of magnitude smaller than $A_{\rm cen}$. This result can be interpreted in terms of a violation of the independence of the outcome of a gravitational experiment from the velocity of the frame in which it is performed, in according with the fact that MOND violates the Strong Equivalence Principle.
Our results are valid not only for EFE in MOND, but also for any other putative exotic modification of gravity yielding a constant and uniform net small extra-acceleration which, thus, can be no larger than  $10^{-14}$ m s$^{-2}$ in the solar system.
It will be important to repeat such a test if and when other teams of astronomers will independently estimate their own corrections to the standard perihelion precessions with different ephemerides.


\end{document}